# A megapixel time-gated SPAD image sensor for 2D and 3D imaging applications

**Kazuhiro Morimoto,**[1,2,*] **Andrei Ardelean,**[1] **Ming-Lo Wu,**[1] **Arin Can Ulku,**[1] **Ivan Michel Antolovic,**[1] **Claudio Bruschini,**[1] and **Edoardo Charbon**[1]

[1]*Advanced Quantum Architecture Laboratory (AQUA), Ecole polytechnique fédérale de Lausanne (EPFL), 2002 Neuchâtel, Switzerland*
[2]*Device Research & Design Department, Canon Inc., 212-8602 Kanagawa, Japan*
*\*Corresponding author: morimoto.kazuhiro@mail.canon*

We present the first 1Mpixel SPAD camera ever reported. The camera features 3.8ns time gating and 24kfps frame rate; it was fabricated in 180nm CIS technology. Two pixels have been designed with a pitch of 9.4μm in 7T and 5.75T configurations, respectively, achieving a maximum fill factor of 13.4%. The maximum PDP is 27%, median DCR 2.0cps, variation in gating length 120ps, position skew 410ps, and rise/fall time <550ps, all FWHM at 3.3V of excess bias. The sensor was used to capture 2D/3D scenes over 2m with an LSB of 5.4mm and a precision better than 7.8mm. Extended dynamic range is demonstrated in dual exposure operation mode. Spatially overlapped multi-object detection is experimentally demonstrated in single-photon time-gated ToF for the first time.

## 1. INTRODUCTION

Time-resolved imaging sensors enable a number of vision techniques, such as time-of-flight imaging, time-resolved Raman spectroscopy, fluorescence lifetime imaging microscopy, super-resolution microscopy, etc. [1-3]. Time-resolved single-photon imaging sensors enable, in addition, quantum vision techniques, such as ghost imaging, sub-shot-noise imaging, quantum LiDAR, quantum distillation, etc. [2,4,5]. Common to these applications is the need for single-photon detection and high timing resolution with low noise and high sensitivity. An important limitation in the majority of the implementations has been the image sensor, usually made of a single pixel or at most a 1kpixel array, thus potentially curtailing acquisition rates. To address these issues, researchers have recently created large-format cameras with a single-photon avalanche diode (SPAD) in each pixel and time gating or time-to-digital converters (TDCs) on chip [6-10]. Though, the crux of a large-format camera remains the pixel pitch and the amount of functionality per pixel. Researchers have thus resorted to 3D integration using backside-illuminated SPADs on the top tier and control/processing/readout electronics on the bottom tier [11-14].

Recently, a novel photon-counting image sensor called quanta image sensor (QIS) has been demonstrated [15-17]. The QIS inherits several advantages of CMOS image sensors such as a potentially small pixel size, high spatial resolution, low dark current, high quantum efficiency and low power consumption. A spatial resolution of up to 1 megapixel with 1.1μm pixels has been reported in a QIS [18,19], enabling low noise and high dynamic range imaging for scientific, space, and security applications. A limitation of the QIS technology, though, is timing resolution. The finite time required for charge transfer in the pixels and sequential scanning readout prevent the QIS from detecting timing information below 1μs. The SPAD, in contrast, enables single-photon detection with a timing resolution of up to few tens of picoseconds owing to the fast avalanche multiplication process. While a TDC-based approach enables precise time stamping of the detected photons in the SPAD array, it is not suited for scaling due to large circuit area and high power dissipation [20-24]. Our time-gating approach, in contrast to [6-9], entails less than 8 transistors, and is promising for scalable photon counting image sensors towards picosecond timing resolution and megapixel spatial resolution. In this paper, we advocate the use of this approach to achieve large-format time-gated SPAD sensors capable of picosecond timing resolution and small pixel pitch.

## 2. IN-PIXEL TIME GATING

Fig. 1(a) shows a simplified schematic of a time-gated SPAD pixel. The SPAD is connected to a quenching transistor to avoid self-sustained avalanche breakdown, and its output signal is selectively fed to an in-pixel memory or counter when a global gate switch is activated. The gate control pulse can be as short as a few nanoseconds. The stored signal is read out through a fast data I/O circuit. Fig. 1(b) illustrates the operation principle of time-gated time-of-flight (ToF) ranging. Laser pulses are repeatedly sent towards the target; the reflected photons are detected at the sensor with a delay of $\Delta t$. Typical time-gating measurement involves consecutive frames with a finely shifted gate window, each of which performs photon counting integrated over $N$ sub-frames. Finer gate scanning improves timing resolution, while sacrificing depth measurement rate or range. From these measurements, a histogram may be derived, as shown in Fig. 1(c). Assuming that the ambient light intensity and the noise of the detector are small enough, sufficient photon counts are obtained when the reflected pulse is captured within the gate window. The resulting photon count profile forms a rectangular distribution with its width corresponding to the gate window width. The delay time $\Delta t$ can be extracted from either rising or falling edge of this profile, while distance $L$ from the detector to the target is estimated by:

$$L = \frac{c \Delta t}{2}, \quad (1)$$

where $c$ is the speed of light.

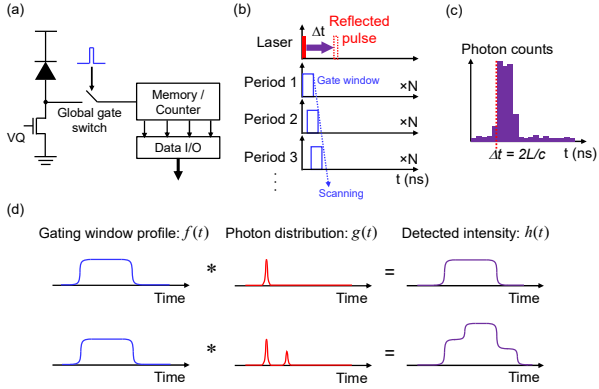

Fig. 1. Conceptual views of time-gated ToF ranging. (a) Pixel circuit architecture of time-gated SPAD sensor. (b) Timing diagram of ToF ranging based on time gate scanning. (c) Expected photon count distribution as a function of gate position. (d) Schematic views of gating window profile, photon distribution and measured intensity over time with single reflective object (top) and double reflective objects (bottom).

More generally, the detected intensity profile $h(t)$ in a given measurement time frame is formulated by the convolution of two functions:

$$h(t) = f(t) * g(t), \quad (2)$$

where $f(t)$ is the gating window profile, and $g(t)$ the photon probability density function. Note that $h(t)$ yields $a$, when integrated from $-\infty$ to $+\infty$, where $a$ is the total detected photon count in the measurement time frame. Fig. 1(d) shows the detected intensity profile of photons captured by the detector characterized by $f(t)$. When the photon probability density function can be approximated by a single Gaussian distribution with a sufficiently small standard deviation, if compared to the gate length, the intensity profile can be expressed as:

$$h(t) \approx f(t) * a \delta(t - \Delta t) = a f(t - \Delta t), \quad (3)$$

where $a$ and $\Delta t$ are the photon count and delay time of the Gaussian peak, respectively. In practice, the detected intensity profile can take more complicated forms. For instance, when the target object is imaged through semi-transparent (semi-reflective) materials such as glass, plastics or liquids, the photon distribution can be expressed as a superposition of multiple Gaussian functions with different peak heights and positions. Assuming again negligible standard deviation, the detected intensity profile is:

$$h(t) \approx f(t) * [\sum_i a_i \delta(t - \Delta t_i)] = \sum_i a_i f(t - \Delta t_i), \quad (4)$$

where $a_i$ and $\Delta t_i$ are the photon count and delay time of the $i$-th Gaussian peak, respectively. Eq. (4) suggests that the multiple reflection results in a superposition of multiple gating window functions, each having different height and delay. An example with two reflective peaks in the photon distribution is shown in the bottom of Fig. 1(d).

Note that when $h(t)$ is measured and $f(t)$ is known, a full profile of $g(t)$ can thus be obtained by deconvolution. In a real situation, $h(t)$ can be distorted by non-ideal effects such as photon-shot noise, ambient light, dark counts, afterpulsing, crosstalk, timing jitter, etc. Those effects can introduce noise in the deconvolution. In ToF ranging, however, the assumption for reflected laser pulses to have negligibly narrow widths is valid in most cases. This assumption simplifies the process of distance calculation, where the time-of-arrival information can be readily extracted by finding the rising or falling edges in the measured intensity profile.

## 3. ARCHITECTURE AND SIMULATION

In this paper, we present the first 1Mpixel camera based on the SPAD pixel described above, with a pitch of 9.4μm. We propose two architectures for the pixel: 7T (pixel A) and 5.75T (pixel B) without and with readout transistor sharing, respectively. The pixels achieve a fill factor of 7.0% and 13.4%, respectively, which can be increased by use of microlenses; both pixels use a dynamic memory to store single-photon events generated by the SPAD. Binary photon counting images are captured and streamed out at 24,000 fps.

Fig. 2(a) and (b) show the schematics and timing diagrams of both pixels (see Supplementary Note S1 and S2 of Supplement 1 for more information). The feedback loop in pixel B prevents any subsequent avalanches within a frame; this is advantageous in very large arrays, since it reduces the current drawn from $V_{OP}$ and thus the power dissipation from that node, which, given the high voltages used, can be significant for pixel counts above 100,000. In our chip, under strong illumination, the current drawn from $V_{OP}$ by pixel A is over 400× that drawn from pixel B (see Supplementary Note S3 of Supplement 1).

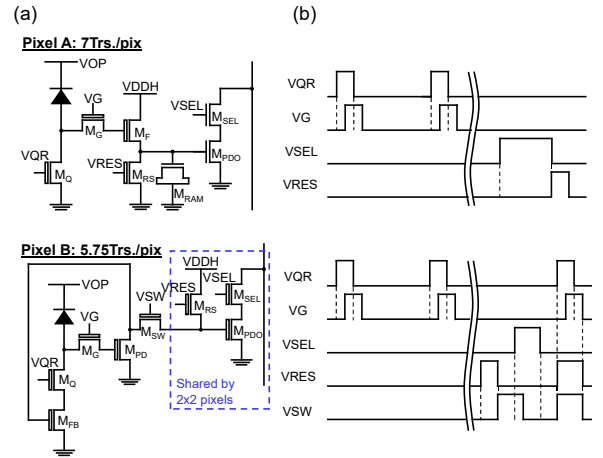

Fig. 2. Schematic views of designed SPAD pixels. (a) Pixel circuit schematics for pixel A and pixel B. (b) Timing charts for pixel circuit operation.

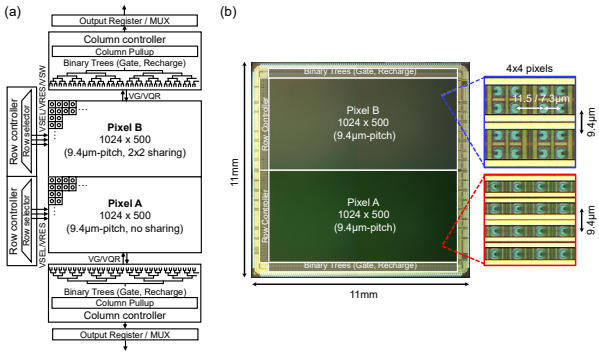

Fig. 3. 1Mpixel time-gated SPAD image sensor architecture. (a) Sensor block diagram. (b) Chip micrograph and magnified views of pixel arrays.

The camera block diagram is shown in Fig. 3(a); it comprises two independent sections of 1024×500 pixels with a dual binary tree controlling the time gate, which reaches a minimum length of 3.8ns and its variation of 120ps (FWHM). Each row is read out in 83ns and stored in a 1024-bit (512-bit) output register for pixel A (pixel B) at the chip bottom, where a multiplexer (MUX) scans it in 128-bit words, which are transferred off-chip by way of a dual parallel bus, thus achieving a frame rate of 24kfps. The micrograph of the image sensor is shown in Fig. 3(b).

## 4. EXPERIMENTAL RESULTS

### A. DCR and PDP

Fig. 4(a) shows the room temperature (RT) cumulative DCR probability distribution of the SPADs throughout the chip, with a median of 0.4cps (pixel A) and 2.0cps (pixel B) at an excess bias of 3.3V. The corresponding DCR per unit drawn active area is 0.065cps/μm$^2$ for pixel A and 0.17cps/μm$^2$ for pixel B. These DCR density metrics are equal to or better than the state-of-the-art SPAD devices [25,26]. Fig. 4(b) shows the measured median DCR as a function of excess bias at room temperature. Fig. 4(c) shows the measured PDP as a function of wavelength. A maximum PDP of 10.5% (pixel A) and 26.7% (pixel B) is reached at 520nm at the same excess bias of 3.3V, while the PDP non-uniformity is better than 1.4% (pixel A) and 3.2% (pixel B) at RT. Lower PDP compared to the previous work based on p-i-n SPAD [25] is caused by the border effect [27]. The border effect is more significant when the active diameter is smaller than 5μm, while the drawn active diameters for pixel A and B are 2.8μm and 3.88 μm, respectively. Fig. 4(d) shows the maximum PDP as a function of the excess bias, whereas the dotted lines are guides for the eye. (see Supplementary Note S4 of Supplement 1 for more detailed analysis).

### B. Time-gating performance

The timing performance of pixel A was characterized in Fig. 5. A 785nm laser pulsed at 25MHz (average power: 5mW, optical pulse width: 80ps, ALS GmbH, Berlin, Germany) illuminates the whole array, while the time gate window is continuously shifted with respect to the laser trigger by steps of 36ps over a range of 10ns. For each gate position, 255 binary frames are acquired and summed in a Kintex™ 7 FPGA (Xilinx Inc., San Jose, CA) to generate an 8-bit image. Fig. 5(a) shows the gate window profiles for 160 pixels uniformly sampled from the bottom-left to the top-right of the pixel array. Broadening of rising and falling edges indicates the non-uniformity of gate signal propagation over the pixel array. Figs. 5(b) and (c) demonstrate the spatial uniformity of gate position and gate length. The gate is activated later in the top side of the array, as gate and recharge signals from the bottom side of the array require more time to propagate. Horizontal skew of the gate position in the top side of the array stems from the asymmetry of power routing where the power and ground are supplied from the left, right and bottom side, but not from the top side of the array. The gate length distribution shows better uniformity than the gate position distribution. Fig. 5(d) shows the histograms of gate position, gate length, rise time and fall time. The gate position skews and variation in gate length were measured at 410ps and 120ps (FWHM), respectively, while an average gate length of 3.8ns was achieved.

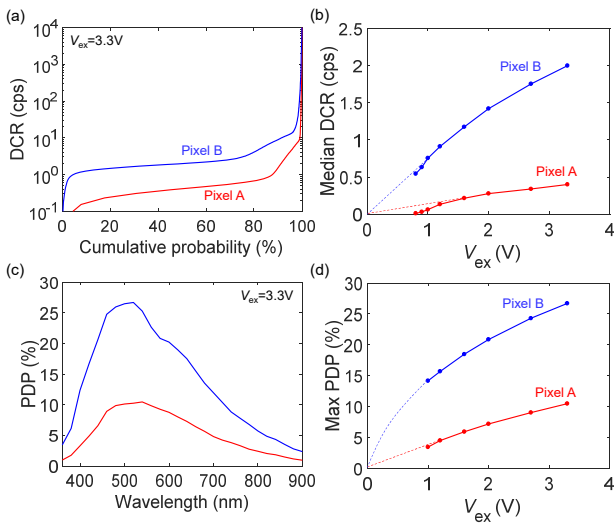

Fig. 4. Measured DCR and PDP for pixel A and B. (a) Room temperature cumulative histogram of DCR at excess bias of 3.3V. (b) Excess bias dependence of median DCR at room temperature. (c) Wavelength dependence of PDP at excess bias of 3.3V. (d) Excess bias dependence of maximum PDP at room temperature.

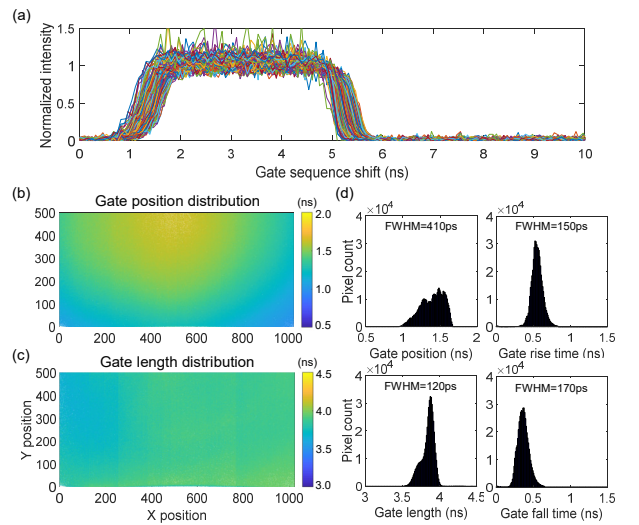

Fig. 5. Measured time-gating performance for pixel A. (a) Gate window profiles for uniformly sampled 160 pixels. (b) Color plot of gate position distribution over 1024×500 pixels. (c) Color plot of gate length distribution over 1024×500 pixels. (d) Histograms for gate position, gate length, rise time and fall time.

## C. 2D imaging

The chip was tested as an intensity image sensor with a standard chart. Fig. 6 shows a 1Mpixel monochrome image obtained at 24kfps with a uniform illumination of 50Lux (indoors). For each half, 16,320 binary images are summed to acquire a 14-bit intensity image. The image contrast for top and bottom half is tuned independently to compensate the difference in the photon detection efficiency (PDE). On the right side of Fig. 6, the magnified images show that the line patterns are well-resolved, up to number 10 in the chart, indicating the spatial resolution of 1000 dots within the horizontal and vertical field of view.

The dynamic range of a 2D image sensor is critical for a wide range of applications. Recently, a method to extend the dynamic range by mixing multiple different exposure times in a single frame has been reported for SPAD-based binary image sensors [28]. Compared to the case with fixed single exposure time for all the binary frames, mixing multiple exposure times results in slower saturation of the output counts when increasing the incident photon flux, giving richer tone for high illumination conditions. Yet, the dynamic range extension based on the interleaved multiple exposure in the SPAD sensor is reported only for a limited sensor resolution of 96×40 pixels. In addition, incident photon count dependencies of output signal, noise and SNR have not yet been systematically compared between single and multiple exposure modes under equalized total exposure conditions.

Fig. 7(a) shows the timing sequences of single and dual exposure modes in time-gated SPAD sensor. The sensor is operated in global shutter mode. Each yellow region in the figure represents a global exposure, followed by sequential readout of a full-resolution binary frame. A set of streamed binary frames is integrated in the FPGA to construct one $N$-bit image. For single exposure mode, the global exposure time is fixed over one $N$-bit frame. For dual exposure mode, short and long global exposures are staggered to form one $N$-bit image. In this experiment, the ratio between short exposure time $\tau_S$ and long exposure time $\tau_L$ for dual exposure mode is set 1 to 8, whereas the exposure time $\tau_M$ for single exposure mode is set $4.5\tau_S$. For systematic comparison of the two operation modes, the maximum photon counts and the total exposure time in a single $N$-bit frame are set equal for two modes; $2\tau_M = \tau_S + \tau_L$.

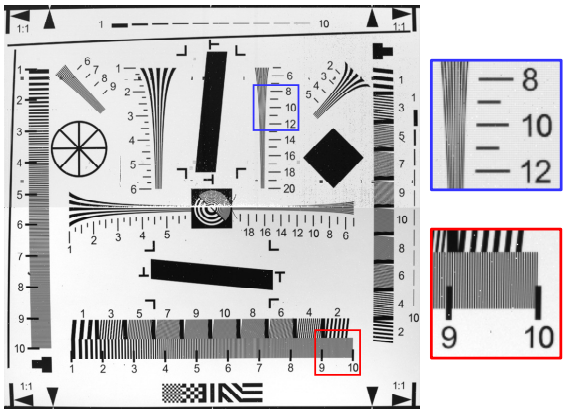

Fig. 6. A 2D intensity image of standard test chart with 1Mpixel resolution. A 14-bit image is obtained by summing 16,320 binary images. Magnified views of two small areas from blue and red squares are shown on the right.

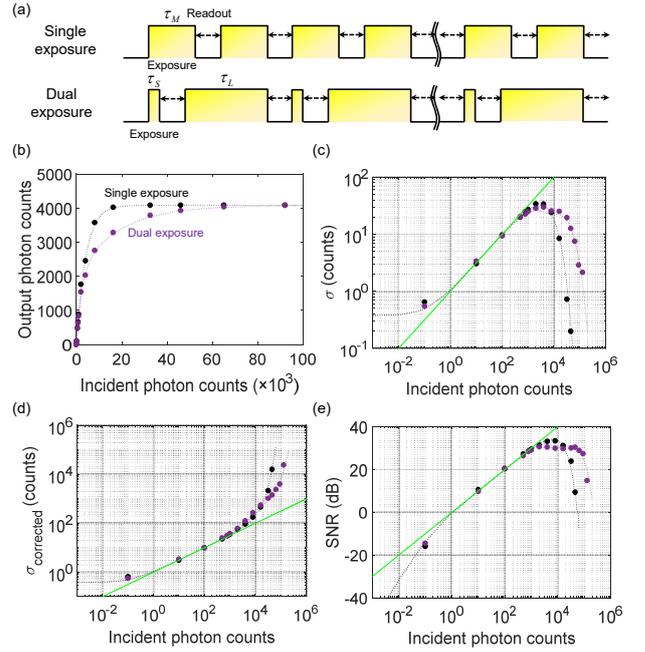

Fig. 7. Conceptual view and measured or simulated results for the dynamic range extension technique. (a) Timing diagrams of single and dual exposure modes. (b) Measured (markers) and fitted (dotted lines) output photon counts as a function of incident photon counts. (c) Measured (markers) and Monte Carlo-simulated (dotted lines) standard deviation. (d) Measured and simulated standard deviation after linearity correction. (e) Measured and simulated SNR. Green lines indicate the photon-shot noise limit.

Fig. 7(b) shows the measured output photon counts as a function of incident photon counts for single and dual exposure modes. 4080 binary frames are summed to form a 12-bit image. Dotted lines are the fitted curves for each mode based on the following equations:

$$N_{out}^S = N_{sat} \times (1 - e^{-\frac{N_{in}}{N_{sat}}}), \quad \textbf{(5)}$$

$$N_{out}^D = \frac{N_{sat}}{2} \times [(1 - e^{-\frac{2\tau_L}{(\tau_L+\tau_S)} \cdot \frac{N_{in}}{N_{sat}}}) + (1 - e^{-\frac{2\tau_S}{(\tau_L+\tau_S)} \cdot \frac{N_{in}}{N_{sat}}})], \quad \textbf{(6)}$$

where $N_{out}^S$ and $N_{out}^D$ are the output counts in the single and dual exposure modes, respectively, $N_{sat}$ is 4080, and $N_{in}$ is the incident photon counts per one $N$-bit frame. The fitted curves are in good agreement with the trends of measured output counts. The output counts of dual exposure mode saturate later than those of single exposure mode, indicating the extended dynamic range.

Fig. 7(c) shows the standard deviation of measured outputs as a function of incident photon counts. Raw output counts of 100 pixels in the center of the array are used to calculate the standard deviation, whereas the photon-shot noise limit is also shown. In the lower incident photon counts, the measured standard deviation is higher than the shot noise limit due to the contribution from DCR non-uniformity. Under intermediate photon counts, the measured standard deviation follows the shot noise limit. For higher incident photon counts, the measured standard deviation is lower than the shot noise limit due to the compression of the output signal when a saturation of 4080 counts is reached [29]. The Monte Carlo simulation results

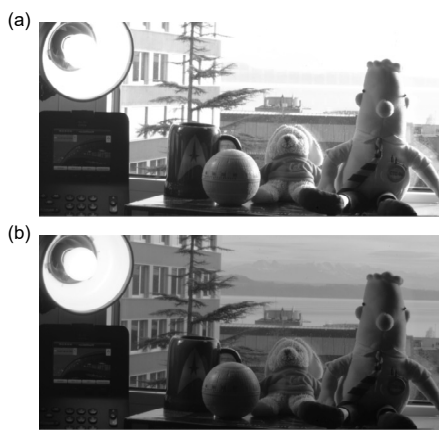

Fig. 8. 2D images of a real-life scene: (a) 18-bit image taken in single exposure mode; (b) 18-bit image taken in dual exposure mode.

for two modes are shown as dotted curves, which are highly consistent with the measured trends.

In real situations, nonlinear output characteristics in Fig. 7(b) has to be corrected to ensure the natural contrast for human eyes. Fig. 7(d) shows the measured standard deviation based on corrected output counts. Similar to the trends in Fig. 7(c), the deviation from photon-shot noise limit is observed in the lower incident photons due to the DCR non-uniformity, and the deviation is suppressed in the intermediate photon counts. For the higher incident photons, the measured trends go above the shot noise limit. The difference with respect to the uncorrected curves in Fig. 7(c) arises from the amplification of photon-shot noise in the linearity correction process. Again, the Monte Carlo simulation (dotted lines) precisely reproduces the measured results for both operation modes. The noise increase in the dual exposure mode is observed later than that of the single exposure mode, which is a direct consequence of dynamic range extension. Fig. 7(e) shows the measured SNR plots for the two exposure modes. A dynamic range of 96.3dB is measured for single exposure and 108.1dB for dual exposure, whereas 11.8dB improvement is demonstrated with equal maximum counts and total exposure time. The highest SNR for single and dual exposure is 33.3dB and 30.5dB, respectively. This implies that the dynamic range and the maximum SNR are in a trade-off relation.

In Fig. 8, the effect of the dynamic range extension is investigated in a real-life scene. In single exposure mode, the background scene is overexposed (Fig. 8(a)), while the gray-scale tone of the scene is clearly visible in dual exposure mode (Fig. 8(b)). The difference of the maximum SNR is too small to be recognized.

**D. 3D imaging**

Fig. 9(a) and (b) show a 2D and a color-coded 3D pictures obtained by illuminating a scene with a 637nm-laser pulsed at 40MHz (average power: 2mW, optical pulse width: 80ps, ALS GmbH, Berlin, Germany) and captured on the half-resolution image sensor (pixel A). The gate window with its width of 3.8ns is shifted from 0.6ns to 13.2ns by steps of 36ps to acquire full photon intensity profiles as a function of the gate position. The distance LSB in this measurement corresponds to 5.4mm. The intensity profile for each pixel is smoothed by taking the moving average over gate positions to suppress the effect of photon-shot noise. The depth information is reconstructed by detecting the rising edge position of the smoothed intensity profile for each pixel, corresponding to the time-of-arrival of the reflected laser pulse. The gate timing skew over the array is compensated by subtracting the background distribution shown in Fig. 5(b) from the measured time-of-arrival distribution. In Fig. 9(b), red color denotes the closer distance from the SPAD camera to the object, whereas blue color corresponds to the farther distance. The maximum depth range for this measurement is set to be 2m, but it can be extended to tens of meters by lowering the laser repetition frequency and increasing the gate step.

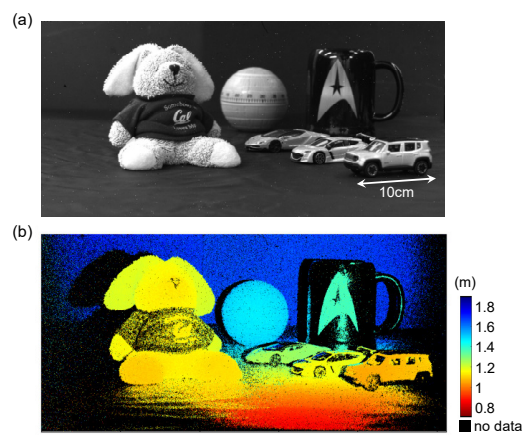

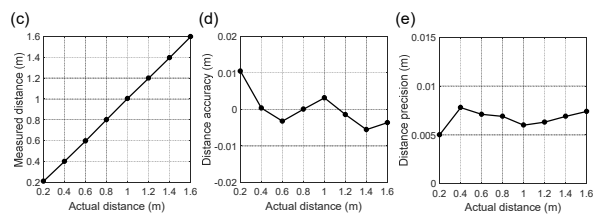

Fig. 9. Measured results for time-gated ToF ranging: (a) real-life 2D intensity image; (b) color-coded 3D image of the same scene obtained with time-gated ToF; (c) measured distance vs actual distance; (d) measured distance accuracy vs actual distance; (e) measured distance precision vs actual distance.

Fig. 9(c) shows the measured distance as a function of the actual object distance. In Fig. 9(c), (d) and (e), a flat object covered with white paper (reflectance around 60%) is used to evaluate the measured distance, accuracy and precision. In Fig. 9(c), the measured distance is extracted by taking the average of the single pixel distance over 20×20 pixels at the center of the array. A very good agreement with the actual distance is observed within the measured range from 0.2 to 1.6m. In Fig. 9(d), the distance accuracy is calculated as the averaged measured distance subtracted by the actual distance. For the measured distance range, the accuracy is always better than 1cm. In Fig. 9(e), distance precision is exploited as a standard deviation of the single pixel distance over 20×20 pixels in the center of the array. The precision is better than 7.8mm (rms) for all the measured points up to 1.6m.

**E. Multi-object 3D imaging**

Compared to indirect ToF [30-32], direct ToF has the advantage that spatially overlapped multiple reflective objects can be imaged individually and accurately. The multi-object detection has been experimentally demonstrated in SPAD-based direct ToF sensors [33,34], where power- and area-consuming TDC circuits and large computational cost for histogramming severely limited the pixel array size of the detector. The multi-object detection has also been demonstrated by either coding temporal illumination or exposure patterns [35,36], which involve a large computational cost to recover the 3D images.
A time-gated ToF sensor provides an alternative, scalable solution by means of compact pixel circuitry and less complicated computation. A CMOS-based time-gating scheme has been adopted for multi-object detection in a 160×120-pixel array [37], where, however, the readout noise limits the lower bound of detectable signal level for each pixel. The readout noise represents a critical issue for scaling the array size because smaller pixel size and larger pixel array size result in the reduced number of reflected photons per pixel, severely limiting SNR. Our gated SPAD pixel enables scalable and readout-noise-free single-photon time gating for multi-object detection.
Fig. 10(a) shows the experimental setup: 510nm-laser beam pulsed at 40MHz (average power: 2mW, optical pulse width: 130ps, PicoQuant GmbH, Berlin, Germany) is spread by a diffuser and used to illuminate a spherical target. The SPAD camera is synchronized with the laser triggering signal, and a transparent plastic plate is inserted between the camera and the object. The distances from the camera to the plastic plate and the object are 0.45m, and 0.75m, respectively. Fig. 10(b) shows 2D intensity images under indoor lighting with and without the plastic plate inserted. Since the plate is almost transparent, no significant difference is observed in the 2D images for those two cases.
The measured time-gating profiles for three representative points (A, B and C) are plotted in Fig. 10(c). Without the plate, the time-gating profiles for point A and B show only a single smoothed rectangular function waveform with its rising edge around gate position 100 (one step of the position corresponding to 36ps). For point C, the photon count stays close to zero over the measured gate position range, indicating no reflective object is detected at this pixel. With the plastic plate, by contrast, the profile at point A shows two-step rising edges around gate positions 40 and 100. Given that the measured profile of photon counts is a convolution of a single smoothed rectangular function and the reflected photon intensity distribution, the two-step profile is a convincing evidence of double reflection from the plastic plate and the spherical object.

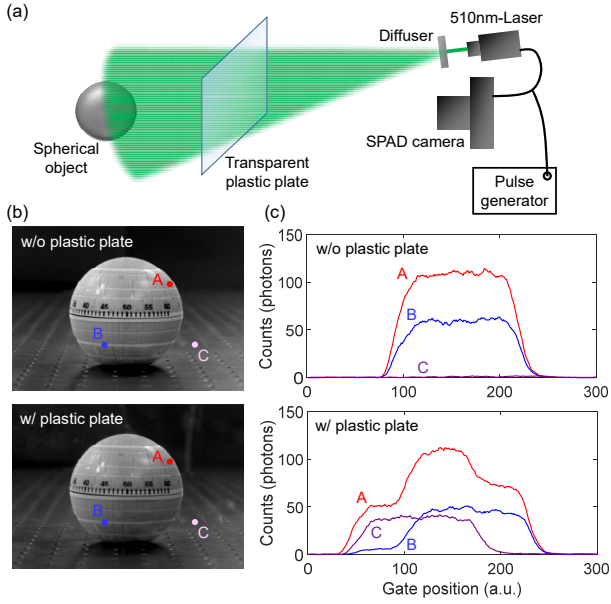

Fig. 10. Experimental setup and measured results for time-gated ToF under multiple reflections. (a) Experimental setup to perform the multi-object detection. (b) Captured 2D images with and without plastic plate. (c) Measured photon count profiles for three different pixels, with and without plastic plate.

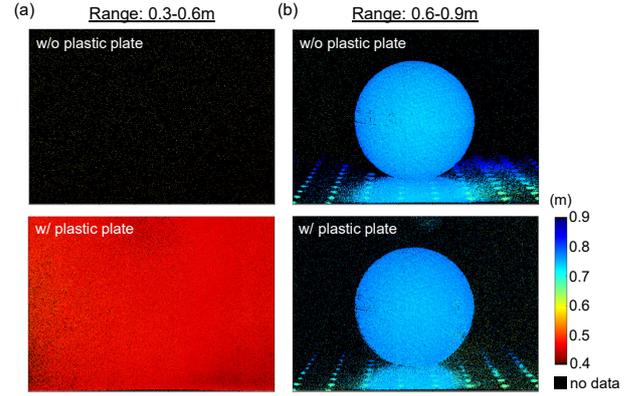

Fig. 11. Reconstructed 3D images in the multi-object detection experiment. (a) 3D images reconstructed based on the distance range of 0.3-0.6m (central 700×500 pixels cropped). Black color indicates that no laser reflection is detected in the measured range. (b) 3D images reconstructed based on the distance range of 0.6-0.9m.

Similar behavior is observed at point B, where the slope of the first rising edge around gate position 40 is milder than that of point A. The profile at point C shows only single rising edge around gate position 40, corresponding to the reflection from the plastic plate. The variation of the slope for the rising edge around gate position 40 between different points is induced by the non-uniform reflection from the surface of the plastic plate.
Fig. 11 shows the reconstructed 3D images based on time-gated ToF. The photon counting profile for each pixel is analyzed to extract the position of rising edges. The rising edge is searched by defining a virtual gate window containing 60 data points of the measured intensity profile. The window is scanned over a whole gate position in non-overlapping fashion, and the existence or non-existence of a rising edge in the virtual window is determined for each scanning position. Fig. 11(a) shows the estimated local distance within the range of 0.3 to 0.6m. Black color pixels represent no object detected within range. Without plastic plate, the vast majority of the pixels shows no detection (black), while the majority of the pixels indicates a reflection at 0.45m (dark red) with the plastic plate, which is consistent with its actual position. Fig. 11(b) shows the estimated distance within the range of 0.6 to 0.9m. For both cases, the distance map of the spherical target object is reconstructed precisely. The measured target object distance is approximately 0.75m, which is also consistent with the actual distance.
The results demonstrate the capability of the time-gated SPAD camera to perform spatially overlapped multi-object detection. Note that the proposed scheme can be applied to the detection of more than two reflection peaks. Finer scanning of the virtual gate window in post-processing enables systematic detection of multiple peaks. The minimum resolvable distance between two neighboring reflective materials is fundamentally limited by the finite rising or falling time of the gate window profile, corresponding to 5-10cm in this SPAD sensor.

## 5. CONCLUSION

In this paper, a 1Mpixel time-gated SPAD image sensor is reported for the first time. In SPAD research, achieving a megapixel SPAD sensor has been considered one of the most important milestones for over a decade [38,39]. The sensor is applied to high dynamic range 2D imaging and high spatio-temporal resolution 3D imaging. To the best of our knowledge, the spatially overlapped multi-object detection with single-photon time-gating scheme has been experimentally demonstrated for the first time. Fig. 12 shows a state-of-the-art comparison of SPAD pixel pitch and array size. The array size of our sensor is the largest, almost 4 times higher than that of the state-of-the-art sensor [10], while the pixel pitch is one of the smallest. A more detailed comparison is summarized in Table 1. Median DCR is the lowest among other works thanks to the optimized process and miniaturized active size. Lower fill factor stems from the front-side illuminated configuration, while it can be further improved, typically by a factor of 2 to 10, by introducing on-chip microlenses [40-42] (see also Supplementary Note S5 of Supplement 1). Owing to its noise and dynamic range performance, the proposed sensor will be useful in a wide variety of industrial applications such as security, automotive, robotic, biomedical, and scientific applications, including quantum imaging and ultra-high-speed imaging.

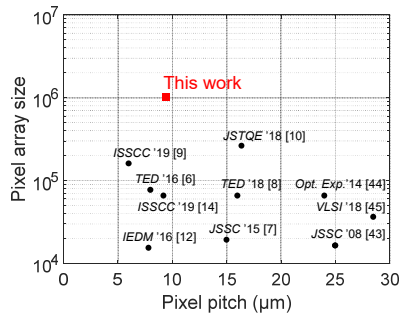

Fig. 12. State-of-the-art comparison of pixel array size and pixel pitch in SPAD sensors.

## FUNDING

This work was funded, in part, by the Swiss National Science Foundation Grant 166289 and by Canon Inc.

## ACKNOWLEDGEMENTS

The authors thank the Swiss National Science Foundation for funding, in part, this research (Grant 166289).

See Supplement 1 for supporting content.

|  | [7] | [6] | [14] | [10] | [9] | This work (pixel A/B) |
|---|---|---|---|---|---|---|
| Process technology | 350nm HV CMOS | 130nm CIS | 40/90nm 3D-BSI | 180nm CMOS | 65nm CIS | **180nm CMOS** |
| Chip size (mm$^2$) | 3.42×3.55 | 3.4×3.1 | - | 9.5×9.6 | - | **11×11** |
| Sensor resolution | 160×120 | 320×240 | 256×256 | 512×512 | 400×400 | **1024×1000** |
| Pixel size (μm) | 15 | 8 | 9.2 | 16.38 | 6 | **9.4** |
| Fill factor (%) | 21 | 26.8 | 51 | 10.5 | 70 | **7.0/13.4** |
| Pixel output bit depth | 5.4b | 1b | 14b | 1b | 1b | **1b** |
| No. of pixel transistors | 8 | 9 | >600 | 11 | 4 | **7/5.75** |
| Median DCR (cps) | 580 ($V_{ex}$=3V) | 47 ($V_{ex}$=1.5V) | 20 ($V_{ex}$=1.5V) | 7.5 ($V_{ex}$=6.5V) | 100 (-) | **0.4/2.0 ($V_{ex}$=3.3V)** |
| Max. PDP (%) | - | 39.5 ($V_{ex}$=1.5V) | 23 ($V_{ex}$=3V) | 50 ($V_{ex}$=6.5V) | - | **10.5/26.7 ($V_{ex}$=3.3V)** |
| Crosstalk (%) | - | - | - | - | - | **0.17/0.39 ($V_{ex}$=3.3V)** |
| Min. gate length (ns) | 0.75 | - | - | 5.75 | - | **3.8** |
| Frame rate (fps) | 486 (5.4b) | 16,000 (1b) | - | 97,700 (1b) | 60 (-) | **24,000 (1b)** |
| Power dissipation (W) | 0.1567 | - | 0.0776 | 0.0267 | - | **0.284/0.535** |

Table 1. State-of-the-art comparison of performance and specifications in large-scale SPAD arrays.

# A megapixel time-gated SPAD image sensor for 2D and 3D imaging applications: supplementary material

KAZUHIRO MORIMOTO,[1,2,*] ANDREI ARDELEAN,[1] MING-LO WU,[1] ARIN CAN ULKU,[1] IVAN MICHEL ANTOLOVIC,[1] CLAUDIO BRUSCHINI,[1] AND EDOARDO CHARBON[1]

[1]*Advanced Quantum Architecture Laboratory (AQUA), Ecole polytechnique fédérale de Lausanne (EPFL), 2002 Neuchâtel, Switzerland*
[2]*Device Research & Design Department, Canon Inc., 212-8602 Kanagawa, Japan*
*\*Corresponding author: morimoto.kazuhiro@mail.canon*

Posted 30[th] December, 2019

This document provides supplementary information to "A megapixel time-gated SPAD image sensor for 2D and 3D imaging applications".

## 1. Supplementary Note S1: pixel circuit operation

The detailed operation principles of pixel A and B are depicted in this section (see Fig. 2 in the main text). Upon detection of a photon, the SPAD generates an avalanche current pulse that is converted to a voltage through quenching transistor $M_Q$, which is controlled by $V_{QR}$. In pixel A, the voltage pulse is transferred to the gate of a transistor acting as memory through a gating transistor $M_G$, controlled by gating signal $V_G$, and follower $M_F$, with a loss of $1V_{TH}$. $M_{RS}$, controlled by $V_{RES}$, is used to reset the dynamic memory, implemented by $M_{RAM}$, to GND. In pixel B, the voltage pulse is transferred to a pulldown transistor $M_{PD}$ via gating transistor $M_G$, which is controlled by gating signal $V_G$. As a result, feedback transistor $M_{FB}$ is set OFF, thus disconnecting the source of $M_Q$ and disabling quenching in the SPAD. The voltage at the drain of $M_{PD}$ is kept near GND for a sufficiently long time until the entire chip is read out and the node is charged again to $V_{DDH}-V_{TH}-V_{DSAT}$ for the next detection, thus turning $M_{FB}$ back ON. Transistor $M_{SW}$, controlled by $V_{SW}$, connects the drain of $M_{PD}$ to the source of $M_{RS}$, which is pre-charged to $V_{DD}-V_{TH}$ via signal $V_{RES}$. In both pixel types, transistor $M_{PDO}$ is used to pull down the entire column when $M_{SEL}$ is turned ON by $V_{SEL}$, i.e. the row is selected in a sequential fashion. The difference is that in pixel B $M_{PDO}$–$M_{SEL}$ are shared among a 2×2-pixel array, while in pixel A the dynamic memory and the readout transistors are thin-oxide devices. All the other transistors are thick-oxide devices to enable operation at 3.3V. $V_{DDH}$, $V_{DD}$, and all the controls (when high) are 3.3V.

## 2. Supplementary Note S2: SPAD device structure

The SPAD, whose cross-section is shown in Fig. S1(a), was implemented as a p+-i-n structure [1], whereas the avalanche region is surrounded by a buried implant at the bottom, to enable uniform field, and a circular guard ring on the sides, to suppress premature edge breakdown, as shown in the cross-section in the figure. All the layers employed in this design are standard in the 180nm CIS process we used in the chip. A simulation of the electric field and of the potential is shown in Fig. S1(b) and (c), demonstrating the location of the avalanche region under the drawn active area, wherever the electric field exceeds the critical field for sustained impact ionization in silicon. The wider depletion region in z-direction with respect to other SPAD structures [2,3] leads to the lower tunneling-induced dark count rate (DCR). The crosstalk was measured in both pixels reaching a mean of 0.17% for pixel A and 0.39% for pixel B. A higher crosstalk in pixel B is expected due to higher proximity to neighbors.

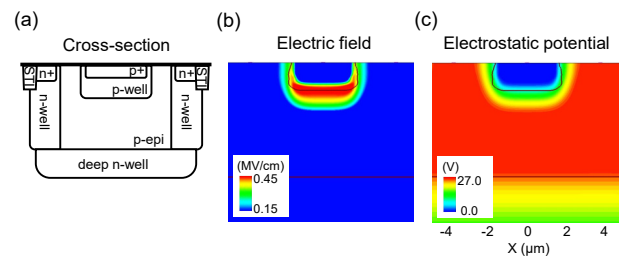

**Fig. S1.** Cross-section and TCAD simulation results of SPAD. (a) Schematic cross-section. (b) Simulated electric field distribution. (c) Simulated Electrostatic potential distribution.

### 3. Supplementary Note S3: power consumption

Fig. S2 shows the measured power consumption of the megapixel SPAD sensor as a function of incident photon flux. Fig. S2(a) is the power consumption for pixel A consisting of $V_{OP}$, 3.3V core, 1.8V core and 3.3V I/O components, respectively. The power consumption at $V_{OP}$ is dominated by avalanche-induced current, and it proportionally increases to the incident photon counts. The power consumption component becomes dominant in the total power consumption under high light condition and reaches 9.118W with SPAD saturation. 3.3V core component is dominated by generation of pixel control signals for VG and VR, and is independent of incident light intensity. 1.8V core originates from pull-up and pull-down of vertical signal lines in the pixel array. The power consumption is proportional to the average photon counts over pixel array, and hence it increases linearly in the lower light and saturates at higher light intensity. 3.3V I/O increases when output binary signal switches frequently between '0' and '1'. It shows the peak at intermediate light intensity where the output binary signal varies randomly. The total power consumption, shown in black curve, exhibits complex behavior due to the mixture of multiple components with different behavior.

Fig. S2(b) is the power consumption for pixel B. Critical difference with respect to pixel A is observed in $V_{OP}$ component. In pixel B, the $V_{OP}$ component saturates at the intermediate incident photon counts and the maximum consumption reaches only 0.021W, approximately 400 times smaller than in pixel A. This stems from the fact that the feedback loop in pixel B closes the recharging path once first photon is detected. This suppresses any extra avalanche multiplication for photons which do not contribute to actual photon counting signal, whereas those extra photons can trigger an avalanche in pixel A due to the different pixel architecture. Another difference is the reversed behavior of 1.8V core caused by the inverted signal output scheme from the pixels. In contrast to pixel A, the power consumption at 1.8V core in pixel B is proportional to the ratio of '0' in output data stream, which shows monotonic decrease as a function of incident photon counts.

### 4. Supplementary Note S4: Temperature dependence of DCR

Fig. S3(a) shows the temperature dependence of DCR for pixel A and B. Based on the temperature dependence of breakdown voltage shown in the inset, $V_{OP}$ is adjusted to keep $V_{ex}$=3.3V over measured temperature range. Median DCR for both pixels shows almost no temperature dependence at T < 10 °C, whereas it increases exponentially at T > 30 °C. For both pixels, the activation energy for T > 30 °C is extracted to be 1.1eV, equivalent to the band gap of silicon. The result indicates that DCR in the majority of pixels is dominated by tunneling at low temperature and by diffusion current at high temperature [4,5].

Fig. S3(b) is the distribution of the activation energy for pixel A, where the horizontal axis corresponds to that of Fig. 4(a) in the main text. Approximately 80% of the pixels in the array show an activation energy of 1.1eV, whereas the remaining 20% with higher DCR exhibits between 1.1eV and 0.55eV. This indicates that those 'hot' pixels have mixed DCR sources from diffusion current and Schockley-Read-Hall (SRH) generation-recombination, which typically lead to the activation energies of 1.1eV and 0.55eV, respectively.

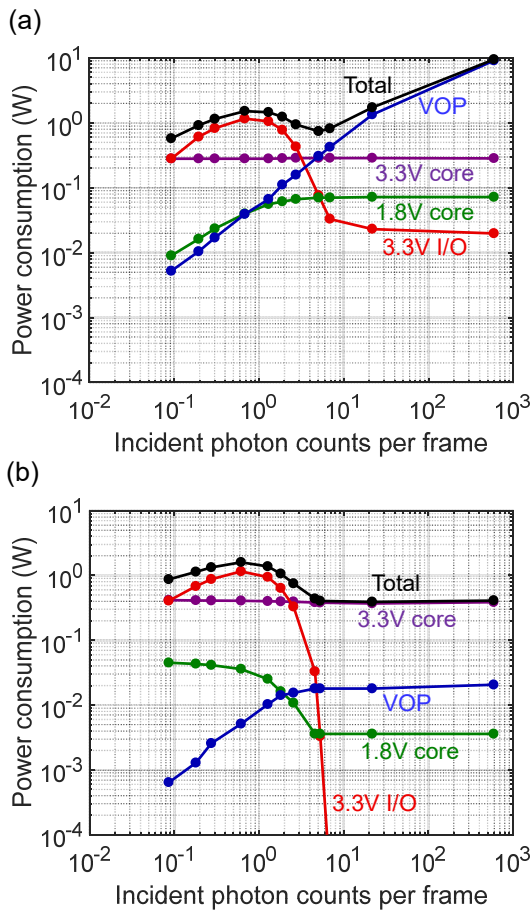

**Fig. S2. Measured power consumption as a function of incident photon counts per frame. (a) Power consumption for pixel A. (b) Power consumption for pixel B.**

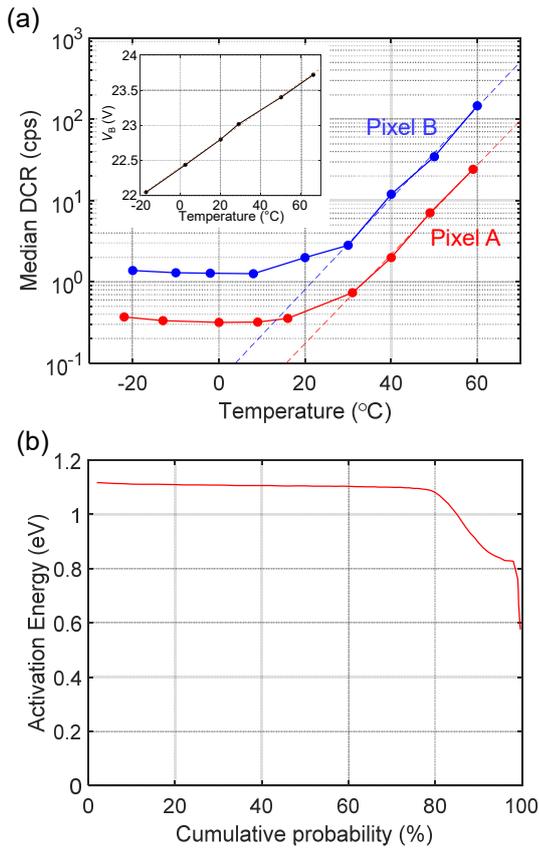

**Fig. S3.** Temperature analysis of DCR. **(a)** Temperature dependence of measured median DCR for pixel A and B. Temperature dependence of breakdown voltage is shown in the inset. **(b)** Activation energy distribution in pixel A, where horizontal axis shows pixel population in ascending order of DCR.

## 5. Supplementary Note S5: Perspectives on pixel pitch reduction

In the CMOS technology node of this paper, we have demonstrated an unprecedented level of miniaturization (for this node) thanks to careful crafting of the pixel and aggressive SPAD size reduction. We believe that further miniaturization is possible and it will be the subject of further research. Preliminary results however suggest that SPAD pitch can still be reduced with relatively minor effects of PDP and DCR. Thus, the bottleneck appears to be the transistor count and sizes. Thus, further pixel miniaturization may involve a more efficient use of the pixel area around the SPAD, including an extension of pixel sharing as proposed in this paper. Moreover, 3D integration and stacking techniques will enable further miniaturization leading to pixels that will continue to scale down to a few microns in pitch. We believe that this evolution will benefit SPAD image sensors with multi-megapixel resolutions in the near future.